\documentclass{ecai}
\usepackage{graphicx}
\usepackage{subfigure}
\usepackage{latexsym}
\usepackage{amsfonts}
\usepackage{amsmath}
\usepackage{algorithm}
\usepackage{bigstrut}
\usepackage{multirow}
\usepackage{multirow}
\usepackage{booktabs}
\usepackage[export]{adjustbox}
\usepackage[noend]{algpseudocode}
\begin{document}

\begin{frontmatter}

\title{A Conditional Denoising Diffusion Probabilistic Model \\ for Radio Interferometric Image Reconstruction}

\author[A]{\fnms{Ruoqi}~\snm{Wang}}
\author[A]{\fnms{Zhuoyang}~\snm{Chen}}
\author[A]{\fnms{Qiong}~\snm{Luo}\thanks{Qiong Luo is the corresponding author. Her other affiliation is the Hong Kong University of Science and Technology.}} 
\author[B]{\fnms{Feng}~\snm{Wang}}

\address[A]{The Hong Kong University of Science and Technology (Guangzhou) \\  \{rwang280, zchen190\}@connect.hkust-gz.edu.cn, luo@ust.hk}
\address[B]{Guangzhou University \\ fengwang@gzhu.edu.cn}

\begin{abstract}

In radio astronomy, signals from radio telescopes are transformed into images of observed celestial objects, or sources. However, these images, called dirty images, contain real sources as well as artifacts due to signal sparsity and other factors.  
Therefore, radio interferometric image reconstruction is performed on dirty images, aiming to produce clean images in which artifacts are reduced and real sources are recovered. So far, existing methods have limited success on recovering faint sources, preserving detailed structures, and eliminating artifacts.
In this paper, we present VIC-DDPM, a Visibility and Image Conditioned Denoising Diffusion Probabilistic Model. Our main idea is to use both the original visibility data in the spectral domain and dirty images in the spatial domain to guide the image generation process with DDPM. This way, we can leverage DDPM to generate fine details and eliminate noise, while utilizing visibility data to separate signals from noise and retaining spatial information in dirty images. We have conducted experiments in comparison with both traditional methods and recent deep learning based approaches.
Our results show that our method significantly improves the resulting images by reducing artifacts, preserving fine details, and recovering dim sources. This advancement further facilitates radio astronomical data analysis tasks on celestial phenomena. Our code is available at https://github.com/RapidsAtHKUST/VIC-DDPM.
\end{abstract}

\end{frontmatter}

\section{Introduction}

\begin{figure}
\centerline{\includegraphics[width=7cm]{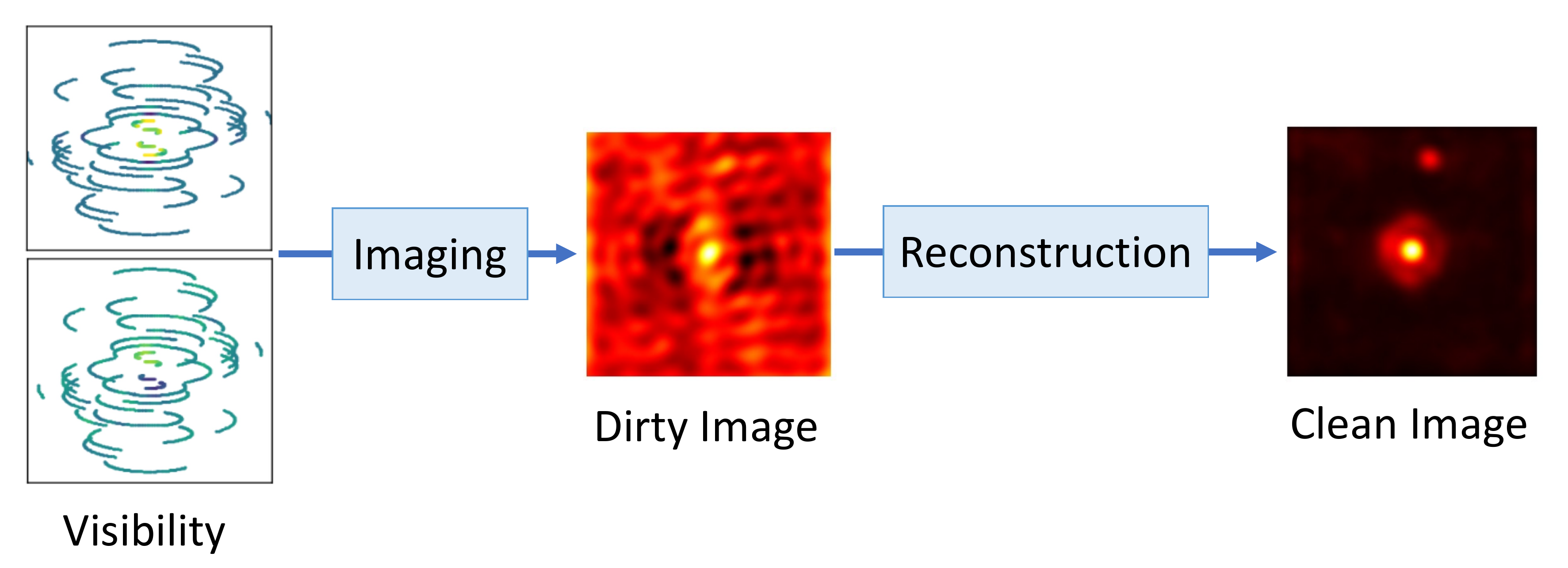}}
\caption{Imaging and Reconstruction}
\label{flow}
\end{figure}

Radio interferometers, or radio telescopes, receive radio waves from astronomical sources and produce interferometric measurement data, called \emph{visibility}.  These visibility data are in the spectral domain, and can be sampled and converted through inverse Fourier transforms into images for subsequent analysis. However, due to data sparsity and other factors, the result images are often dominated by artifacts \cite{schmidt2022deep}. As such, these \emph{dirty images} must be reconstructed before they can be used in scientific analysis. Figure \ref{flow} illustrates the workflow of imaging and reconstruction.
In this paper, we propose an image reconstruction method, aiming to effectively recover real sources and eliminate artifacts in dirty images.

Reconstruction of interferometric images is a challenging problem. Its effectiveness heavily depends on suitable priors to select from a vast array of potentially valid images in an infinite solution space \cite{wu2022neural}. For example, the classic CLEAN method \cite{hogbom1974aperture} uses human knowledge as priors and assumes that the sky is composed of point-like sources \cite{connor2022deep}. As real celestial objects can be either point or extended sources, this assumption limits the quality of reconstructed images \cite{wu2022neural}. Therefore, recent deep-learning based approaches \cite{morningstar2018analyzing,morningstar2019data,wu2022neural, schmidt2022deep} learn data-driven priors from training datasets. These methods successfully predict general structures of main objects in observation areas, but they still struggle in recovering faint sources, preserving details, and eliminating artifacts.

To perform high-quality radio interferometry reconstruction, we propose VIC-DDPM, a Visibility and Image Conditioned Denoising Diffusion Probabilistic Model (DDPM).
We adopt a DDPM \cite{ho2020denoising, sohl2015deep} because it can generate images with fine details. We use both the original visibility data in the spectral domain and dirty images in the spatial domain to guide the clean image generation with DDPM. This way, our model can leverage the complementary strengths of both domains: Visibility data offers frequency information where noise can be effectively separated from signals \cite{boulch2016deep}, aiding in identifying artifacts; dirty images provide spatial information, helping recover dim sources and object structures. Consequently, our model removes artifacts as well as restores fine details, including details in main structures and surrounding dim sources. To the best of our knowledge, VIC-DDPM is the first work that adapts diffusion models to radio interferometric image reconstruction. 

In short, our main contributions are as follows:
\begin{itemize}
    \item We present a DDPM based generative model with two conditioning mechanisms, for dirty image and visibility data, respectively, in radio interferometric image reconstruction.
    \item We show that data-driven priors in interferometric datasets can be effectively learned by incorporating both visibility and image data.
\end{itemize}

We have conducted experiments to evaluate VIC-DDPM in comparison with both popular traditional methods and a few recent state-of-the-art deep-learning based models. Our results show that our method significantly improves the output images. In particular, VIC-DDPM removes artifacts more thoroughly, reconstructs structural details better, and recovers surrounding dim sources. This advancement facilitates a more comprehensive and accurate analysis of radio astronomical data, enabling researchers to better understand complex celestial phenomena.

The remainder of this paper is organized as follows. In section \ref{Related Work}, we describe the background of radio interferometric imaging and image reconstruction, related reconstruction algorithms, and basic denoising diffusion probabilistic models. In section \ref{Methods}, we present the details of VIC-DDPM, including the two conditioning factors -- dirty image and visibility. In section \ref{Experiments and Results}, we describe experimental settings and present the results with analysis.

\section{Background and Related Work}
\label{Related Work}
\subsection{Interferometric Imaging}
In radio astronomy, observational data are gathered from radio telescopes, or \emph{interferometers}, which detect radio signals from remote celestial objects. These interferometers utilize aperture synthesis \cite{burke2019introduction} to process their captured signals, generating \emph{visibility} data through Fourier transforms of the sky's brightness distribution \cite{liu2022efficient}. Then, the \emph{imaging} process constructs images of the observed sky from the visibility data for subsequent analysis \cite{thompson2017interferometry} .

Specifically, as the complex value of the visibility of an astronomical object is equivalent to the Fourier transform of the object's intensity distribution function, we can apply an inverse Fourier transform to convert the $(u, v)$ coordinates in the Fourier domain into $(l, m)$ in the image domain \cite{wu2022neural}:
\begin{eqnarray}
I(l, m) & = & \int_{u} \int_{v} e^{2 \pi i(u l+v m)} V(u, v) d u d v
\end{eqnarray}
where $V(u, v)$ denotes the complex visibility in the Fourier space, and $I(l, m)$ denotes the intensity distribution in the image. 

\subsection{Interferometric Image Reconstruction}
Due to the limited number of radio telescopes and their non-uniform distribution on the ground,  visibility data is often sparsely and irregularly sampled, leading to an incomplete representation of the true sky. When the inverse Fourier transform is applied to this sparse data, the resulting image, called a \emph{dirty image}, is dominated by artifacts \cite{schmidt2022deep}. For better applicability in scientific analysis, dirty images must go through further reconstruction to enhance the clarity of sources and reduce artifacts.

Interferometric image reconstruction is an ill-posed problem because there are infinite  solution images that fit the observed visibility \cite{sun2021deep}. Traditional reconstruction methods \cite{ables1974maximum, hogbom1974aperture} utilize human prior knowledge to reduce the solution space.
In particular, the maximum entropy method (MEM) \cite{ables1974maximum} mathematically determines the solution that best agrees with the available sparse measurement, whereas 
the CLEAN algorithm \cite{hogbom1974aperture} iteratively operates on the dirty image to distinguish real structures from sidelobe disturbances. For the past decades, CLEAN \cite{hogbom1974aperture} has served as the main-stream radio interferometric reconstruction method; however, its resulting images have limitations \cite{wu2022neural} due to its assumption of all point-like sources in the sky \cite{connor2022deep}. 

Recently, a flurry of deep-learning based data processing methods have been proposed in radio astronomy and computational imaging. Morningstar et al. \cite{morningstar2018analyzing,morningstar2019data} used neural networks in radio interferometry and can generate reproducible results \cite{schmidt2022deep}. 
Sun et al. presented a variational depth probability imaging method for quantifying reconstruction uncertainty, which can efficiently sample posteriori distributions \cite{sun2021deep}. Schmidt et al. \cite{schmidt2022deep} used radionets with a convolutional neural network structure to generate reproducible clean radio images on simulated data. Wu et al. \cite{wu2022neural} proposed to perform sparse-to-dense inpainting in the spectral domain with coordinate-based neural fields, outperforming other established interferometry techniques.

We experimented several of these state-of-the art deep learning based methods \cite{wu2022neural, schmidt2022deep, ronneberger2015u} on galaxy images each of which contains a number of astronomical sources, and found that these methods reconstructed the main structure of the most obvious object in the image. However, some details and surrounding dim sources were missing, and considerable amounts of noise remained in the reconstructed image. 
Therefore, we turn to diffusion models, which could potentially address these limitations, since they are able to generate diverse high-resolution images with high semantic fidelity of conditioning \cite{takagi2022high}. To the best of our knowledge, no prior studies have used diffusion models to reconstruct dirty images in radio interferometry.

\begin{figure*}
\centerline{\includegraphics[height=6cm]{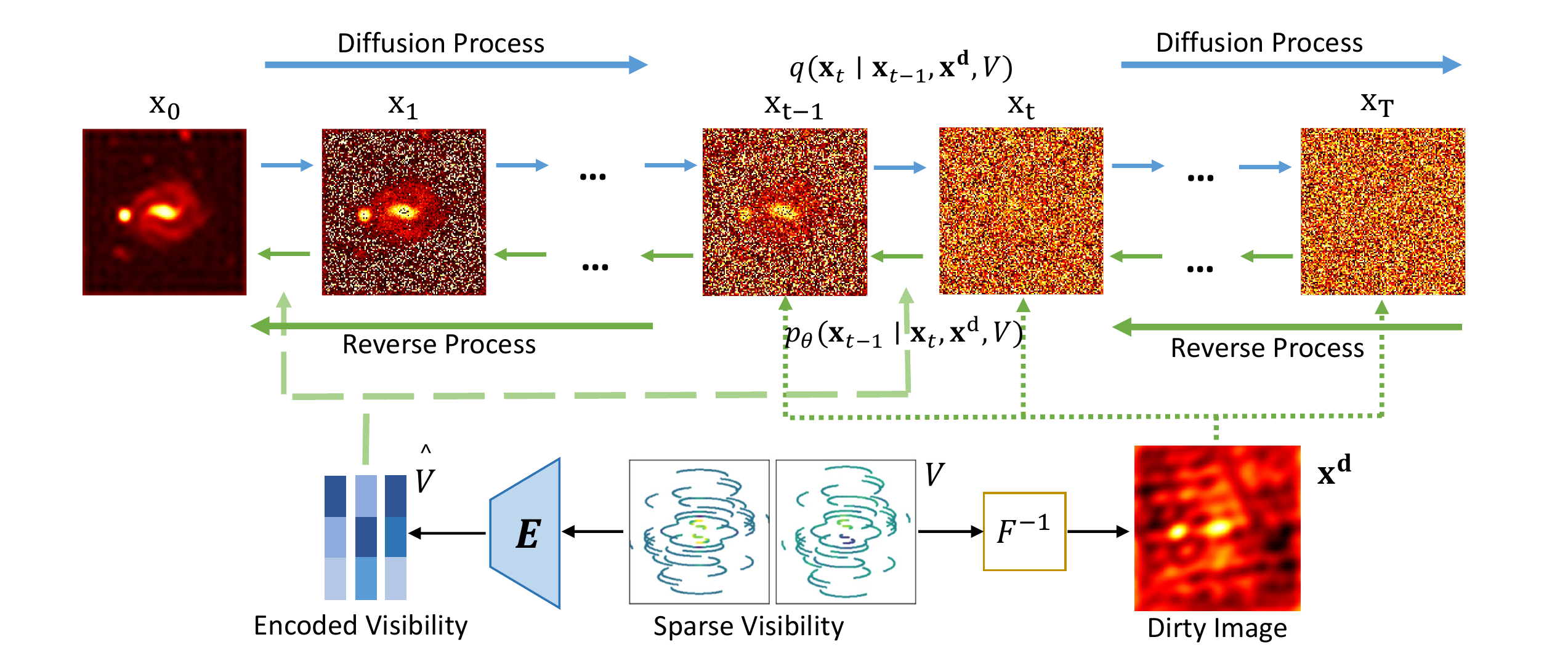}}
\caption{Overview of the Proposed VIC-DDPM. The diffusion process starts from $\mathbf{x_0}$ and Gaussian noise is gradually added to $\mathbf{x_{t-1}}$ to obtain $\mathbf{x_{t}}$ until $t=T$. The reverse process starts from random noise and generates $\mathbf{x_{t-1}}$ from $\mathbf{x_{t}}$. Visibility data and dirty images are added to the reverse process to condition the clean image generation. Details are described in Section \ref{formulation}.}
\label{Overview}
\end{figure*}

\subsection{DDPM}
Denoising Diffusion Probabilistic Models (DDPMs) are a class of deep generative models that learn to generate realistic samples from a given data distribution. 
The core idea behind DDPMs, introduced by Sohl-Dickstein et al. \cite{sohl2015deep} and further developed by Ho et al. \cite{ho2020denoising}, is to reverse a denoising process that transforms the data distribution into Gaussian noise. The diffusion process can be thought of as a series of noise-corrupted versions of the original data, where each step progressively adds more noise to the data until the result becomes Gaussian noise \cite{sohl2015deep}. The model then learns to denoise the data by predicting the preceding, less noisy version of the data at each step, given the current noisy version. To generate new samples, the model starts from Gaussian noise and simulates the reverse diffusion process by iteratively denoising the data at each step \cite{ho2020denoising}. As the model has learned to denoise data at every step of the process, the generated samples become progressively more realistic and structured.

Diffusion models show great generative capabilities to produce fine levels of details \cite{croitoru2023diffusion}. Current diffusion architectures are well designed to work with image data in the spatial domain. When their underlying neural backbone is implemented as a U-Net, the generative capacity of these models shifts from the natural fitting of inductive biases to image-like data \cite{dhariwal2021diffusion,ho2020denoising,ronneberger2015u,song2020score,rombach2022high}. Latent diffusion models further introduce cross-attention layers into the U-Net model architecture, making diffusion models more powerful and flexible with general conditioning input \cite{takagi2022high}. In our work, we propose to apply cross-attention with visibility data and dirty image for radio interferometric image reconstruction.

\section{Our Method}
\label{Methods}
In this section, we present our Visibility and Dirty Image Conditional DDPM (VIC-DDPM) for radio interferometric image reconstruction. We first give an overview of our method and formulate the VIC-DDPM. Then we present the details of visibility and dirty image conditioning, respectively.

\subsection{Formulation of  \space VIC-DDPM}
\label{formulation}
Suppose $\mathbf{x} \in \mathbb{R}^{n}$ is an image of real sky's brightness distribution and $V \in \mathbb{R}^{m}$ is the sampled visibility. They follow the forward model:
\begin{eqnarray}
V=\mathbf{P}_{\Omega} F(\mathbf{x})+\epsilon^\prime
\end{eqnarray}
where $F$ refers to the Fourier transformation, $\mathbf{P}_{\Omega} \in \mathbb{R}^{m \times n}$ is an under-sampling matrix with $\Omega$ denoting the sample pattern produced by the sampling function of the telescope, and $\epsilon^\prime$ denotes noise. 

The task of radio interferometric image reconstruction can be simplified to estimating the posterior distribution $q\left(\mathbf{x} \mid V\right)$. Since $\mathbf{x^d} = F^{-1}\left(V\right)$, the task can then be converted to estimate $q\left(\mathbf{x} \mid \mathbf{x^d}, V\right)$, leveraging the complementary advantages of visibility observations and dirty images. Then we define our diffusion model in the following form:
 \begin{eqnarray}
p_{\theta}\left(\mathbf{x}_{0} \mid \mathbf{x^d}, V\right): =  \int p_{\theta}\left(\mathbf{x}_{0: T} \mid \mathbf{x^d}, V\right) \mathrm{d} \mathbf{x}_{1: T}.
\end{eqnarray}

The reverse process $p_{\theta}\left(\mathbf{x}_{0: T} \mid \mathbf{x^d}, V\right)$ is defined as:
\begin{eqnarray}
p_{\theta}\left(\mathbf{x}_{0: T} \mid \mathbf{x^d}, V\right)=p\left(\mathbf{x}_{T}\right) \prod_{t=1}^{T} p_{\theta}\left(\mathbf{x}_{t-1} \mid \mathbf{x}_{t}, \mathbf{x^d}, V\right),\\
p_{\theta}\left(\mathbf{x}_{t-1} \mid \mathbf{x}_{t}, \mathbf{x^d}, V\right) = \mathcal{N}\left(\mathbf{x}_{t-1} ; \boldsymbol{\mu}_{\theta}\left(\mathbf{x}_{t}, t, \mathbf{x^d}, V\right), \sigma_t ^2 \mathbf{I}\right).
\end{eqnarray}
Similar to the original DDPM \cite{ho2020denoising}, we fix the diffusion process to a Markov chain, defined as:
\begin{eqnarray}
q\left(\mathbf{x}_{1: T} \mid \mathbf{x}_0, \mathbf{x^d}, V\right) : = \prod_{t = 1}^{T} q\left(\mathbf{x}_t \mid \mathbf{x}_{t-1}, \mathbf{x^d}, V\right),\\
q\left(\mathbf{x}_t \mid \mathbf{x}_{t-1}, \mathbf{x^d}, V\right) : = \mathcal{N}\left(\mathbf{x}_t; \alpha_{t} \mathbf{x}_{t-1}, \beta_{t}^{2} \mathbf{I}\right).
\end{eqnarray}
The diffusion process gradually adds Gaussian noise to data based on scheduled $\alpha_1,...,\alpha_t$ and $\beta_1,...,\beta_t$. 

Different from the original DDPM, we follow most recent work \cite{dhariwal2021diffusion, xie2022measurement} to set the $\alpha_t$ and $\beta_t$ schedule.

Let $\bar{\alpha}_{t} = \prod_{i = i}^{t} \alpha_{i}, \bar{\beta}_{t}^{2} = \sum_{i = 1}^{t} \frac{\bar{\alpha}_{t}^{2}}{\bar{\alpha}_{i}^{2}} \beta_{i}^{2}, \bar{\alpha}_{0} = 1, \bar{\beta}_{0}=0$:

\begin{eqnarray}
q\left(\mathbf{x}_{t} \mid \mathbf{x}_0, \mathbf{x^d}, V\right) &= \mathcal{N}\left(\mathbf{x}_t; \bar{\alpha}_{t} \mathbf{x}_0, \bar{\beta}_{t}^{2} \mathbf{I}\right),
\end{eqnarray}
\begin{eqnarray}
q\left(\mathbf{x}_{t-1} \mid \mathbf{x}_t, \mathbf{x}_0, \mathbf{x^d}, V\right) &= \mathcal{N}\left(\mathbf{x}_{t-1}; \tilde{\boldsymbol{\mu}}_{t}\left(\mathbf{x}_t, \mathbf{x}_0\right), \bar{\beta}_{t}^{2} \mathbf{I}\right), \label{eq}
\end{eqnarray}
where 
\begin{equation}
\begin{aligned}
\tilde{\boldsymbol{\mu}}_{t}\left(\mathbf{x}_t, \mathbf{x}_0\right)  =  \frac{\alpha_{t} \bar{\beta}_{t-1}^{2}}{\bar{\beta}_{t}^{2}} \mathbf{x}_t +\frac{\bar{\alpha}_{t-1} \beta_{t}^{2}}{\bar{\beta}_{t}^{2}} \mathbf{x}_0, \text{and } \tilde{\beta}_{t} =  \frac{\beta_{t} \bar{\beta}_{t-1}}{\bar{\beta}_{t}}.
\end{aligned}
\end{equation}

With our schedule setting of $\alpha_t$, we have $\bar{\alpha}_T \approx 0$.

Then, we perform the training of $p_{\theta}\left(\mathbf{x}_{0} \mid \mathbf{x}_{t-1}, \mathbf{x^d}, V\right)$ by optimizing the variational bound on the negative log likelihood, similar to DDPM \cite{ho2020denoising}:
\begin{equation}
\begin{aligned}
&\mathbb{E}\left[-\log p_{\theta}\left(\mathbf{x}_{0} \mid \mathbf{x^d}, V\right)\right]  \\ 
& \leq \mathbb{E}_{q}\left[-\log \frac{p_{\theta}\left(\mathbf{x}_{0: T} \mid \mathbf{x^d}, V\right)}{q\left(\mathbf{x}_{1: T} \mid \mathbf{x}_{0}, \mathbf{x^d}, V\right)}\right] \\ 
& = \mathbb{E}_{q}\Bigg[-\log p\left(\mathbf{x}_{T} \mid  \mathbf{x^d}, V\right)\\ 
& \qquad \quad -\sum_{t \geq 1} \log \frac{p_{\theta}\left(\mathbf{x}_{t-1} \mid \mathbf{x}_{t},  \mathbf{x^d}, V\right)}{q\left(\mathbf{x}_{t} \mid \mathbf{x}_{t-1},  \mathbf{x^d}, V\right)}\Bigg] \\ 
&= : L.
\end{aligned}
\end{equation}

We choose the parameterization:
\begin{equation}
\begin{aligned}
\boldsymbol{\mu}_{\theta}\left(\mathbf{x}_{t}, t, \mathbf{x^d}, V\right) 
& = \frac{1}{\alpha_{t}}\left(\mathbf{x}_{t}-\frac{\beta_{t}^2}{\bar{\beta_{t}}} \boldsymbol{\epsilon}_{\theta}\left(\mathbf{x}_{t}, t, \mathbf{x^d}, V\right)\right) \label{para}
\end{aligned}
\end{equation}
where $\boldsymbol{\epsilon}_{\theta}$ is a function approximator for predicting $\boldsymbol{\epsilon}$ from $\mathbf{x}_{t}$. If $t=0$, we set $\mathbf{z}_t = 0$; otherwise, we sample $\mathbf{x} \sim p_{\theta}\left(\mathbf{x}_{t-1} \mid \mathbf{x}_{t}, \mathbf{x^d}, V\right)$ by computing
$\mathbf{x}_{t-1} = \frac{1}{\alpha_{t}}\left(\mathbf{x}_{t}-\frac{\beta_{t}^2}{\bar{\beta_{t}}} \boldsymbol{\epsilon}_{\theta}\left(\mathbf{x}_{t}, t, \mathbf{x^d}, V\right)\right) + \sigma_t \mathbf{z}_t$, where $\mathbf{z}_t \sim \mathcal{N}(0, \mathbf{I})$.

Then, based on (Eq. \ref{eq}), and with the parameterization (Eq. \ref{para}), we simplify $L$ to:
\begin{eqnarray}
L_{simple}(\theta) = \mathbb{E}_{\mathbf{x}_{0}, t,  \boldsymbol{\epsilon}}\left\|\boldsymbol{\epsilon}-\boldsymbol{\epsilon}_{\theta}\left(\bar{\alpha}_{t} \mathbf{x}_{0}+\bar{\beta}_{t} \boldsymbol{\epsilon}, t, \mathbf{x^d}, V\right)\right\|_{2}^{2}\label{loss}
\end{eqnarray}
where $\boldsymbol{\epsilon} \sim \mathcal{N}\left(\mathbf{0}, \mathbf{I}\right)$ and $t \sim \text{Uniform}({0 , ... , T})$.

In summary, Figure \ref{Overview} illustrates the diffusion process and reverse process in VIC-DDPM. Algorithm \ref{al1} presents the diffusion process using the loss function (Eq. \ref{loss}). Algorithm \ref{al2} shows the sampling procedure.

\begin{algorithm}[t]
\caption{Training}
\label{al1}
\begin{algorithmic}[1]
  \Repeat
    \State $\{\mathbf{x}\} \sim q \left(\mathbf{x}\right)$, obtain $V$, $\mathbf{x^d} = F^{-1}\left(V\right)$
    \State $\boldsymbol{\epsilon} \sim \mathcal{N}(0, \mathbf{I})$
    \State $t \sim \text{Uniform}({0 , ... , T})$
    \State $\mathbf{x}_t = \bar{\alpha}_t \mathbf{x}_0 + \bar{\beta_t} \boldsymbol{\epsilon}$
    \State Compute the loss $L(\theta) = \|\boldsymbol{\epsilon}-\boldsymbol{\epsilon}_{\theta}\left(\mathbf{x}_t, t,\mathbf{x^d},V\right)\|_{2}^{2}$
    \State Update $\theta$ with gradient descent on $L(\theta)$
  \Until{convergence}

\end{algorithmic}
\end{algorithm}

\begin{algorithm}[t]
\caption{Sampling}
\label{al2}
\begin{algorithmic}[1]
  \State Given $V$, $\mathbf{x^d} = F^{-1}\left(V\right)$
  \For{$t = T, ..., 1$, }
    \State $\mathbf{z}_t \sim \mathcal{N}(0, \mathbf{I})$ if $t>1$, else $\mathbf{z}_t = 0$
    \State $\mathbf{x}_{t-1} = \frac{1}{\alpha_{t}}\left(\mathbf{x}_{t}-\frac{\beta_{t}^2}{\bar{\beta_{t}}} \boldsymbol{\epsilon}_{\theta}\left(\mathbf{x}_{t}, t, \mathbf{x^d}, V\right)\right) + \sigma_t \mathbf{z}_t$
  \EndFor
  \State \textbf{return} $\mathbf{x}_0$

\end{algorithmic}
\end{algorithm}

\subsection{Details of Conditioning}
\label{Conditioning Details}
\begin{figure}
\centerline{\includegraphics[width=8cm]{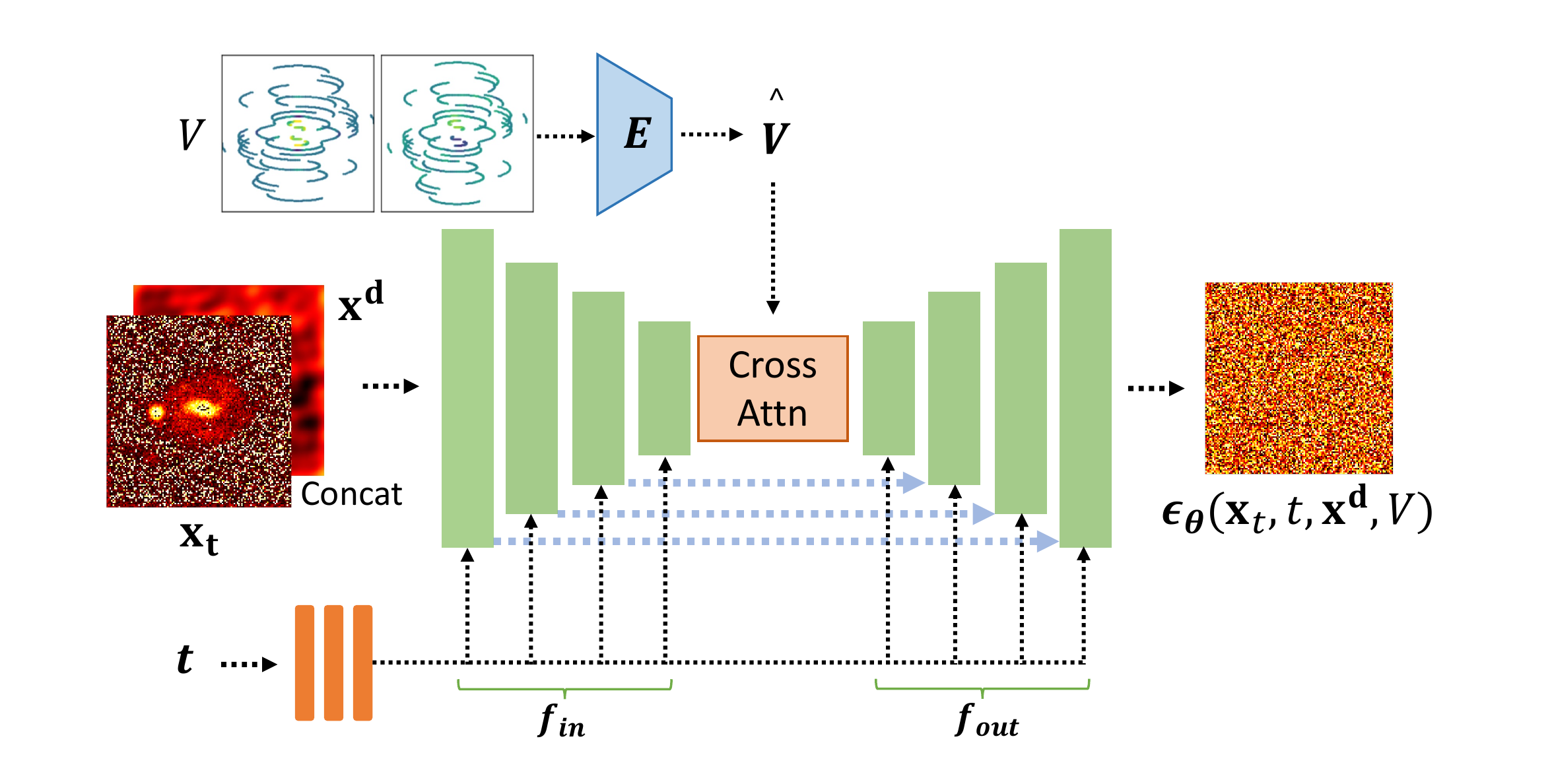}}
\caption{\textbf{The U-Net architecture in our method to represent $\boldsymbol{\epsilon}_{\theta}\left(\mathbf{x}_{t}, t, \mathbf{x^d}, V\right)$.}}
\label{unet}
\end{figure}

We use the U-Net \cite{ronneberger2015u} architecture with ResNet \cite{he2016deep} blocks and attention layers \cite{vaswani2017attention} as our backbone neural network to represent  $\boldsymbol{\epsilon}_{\theta}\left(\mathbf{x}_{t}, t, \mathbf{x^d}, V\right)$.  Figure \ref{unet} gives an overview of the network.

\subsubsection{Dirty Image Conditioning} 
 As shown on the left in Figure \ref{unet}, we concatenate the dirty image $x_d$ with the input $x_t$ of the U-Net, to leverage the information provided by the dirty image. The concatenation is done by each channel of the image. This way, we enable the U-Net to extract features from both the input data and the dirty image, thereby incorporating prior information from the dirty image into the reconstruction process. 

\subsubsection{Sparse Visibility Encoding} 
Visibility data is sparse, so we apply encoding to increase the information density and then incorporate the encoded visibility $\hat{V}$ into our network, as shown in the top left of Figure \ref{unet}. 

\begin{figure}[b]
\centerline{\includegraphics[width=7cm]{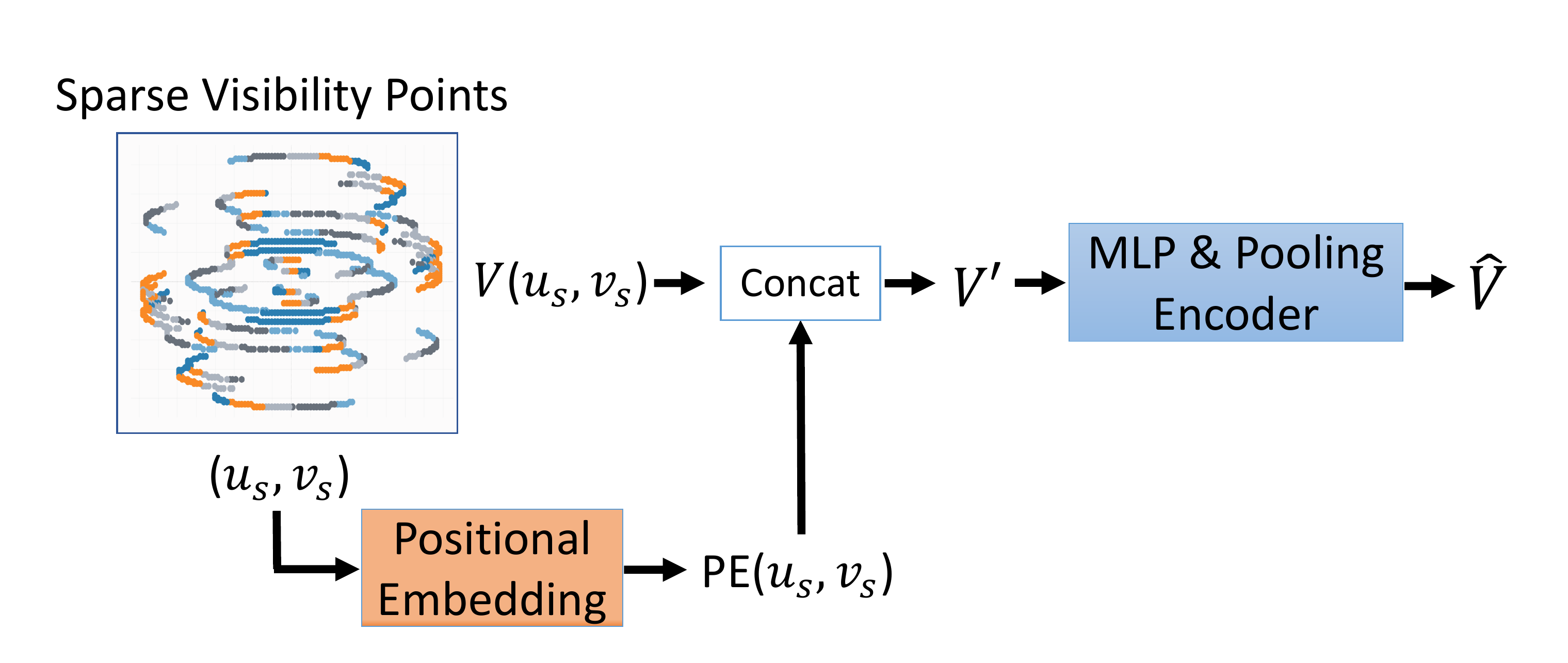}}
\caption{\textbf{Sparse Visibility Encoding.}}
\label{visenco}
\end{figure}

Figure \ref{visenco} illustrates our encoding method. As shown on the left in Figure \ref{visenco}, the sparsely sampled visibility data is in the form of $\left\{u_{s}, v_{s}, {V}\left(u_{s}, v_{s}\right)\right\}$, where $(u_{s}, v_{s})$ are the coordinates at which a measurement is sampled and ${V}\left(u_{s}, v_{s}\right)$ is the complex value of the sample.

To integrate each visibility value and its corresponding position, we first encode each sample individually using positional embedding ($\text{PE}(u_s,v_s)$ in Figure \ref{visenco}). Specifically, we first encode the positional information of a sample point using a NeRF-style \cite{mildenhall2021nerf} embedding method (axis-aligned, power-of-two frequency sinusoidal encoding) \cite{wu2022neural}. 
It facilitates the following Multi-Layer-Perception (MLP) to learn high frequency information \cite{tancik2020fourier, wu2022neural}. 
After that, the positional embedding $\text{PE}(u_s,v_s)$ and the complex value of visibility ${V}\left(u_{s}, v_{s}\right)$ are concatenated to form visibility tokens $V^{\prime}$.

Denote the sparse visibility tokens as $V^{\prime}=\left[v^{\prime}_{1} ; v^{\prime}_{2} ; v^{\prime}_{3} ; \ldots v^{\prime}_n\right]$, and $V^{\prime} \in \mathbb{R}^{n \times d}$, where $n$ is the number of sampled measurement points and $d$ is the number of dimensions of visibility tokens. Furthermore, the sampled points are divided into $m$ groups by their rotation angles. Then, we apply a linear mapping layer $M$ and a mean pooling for encoding, as shown on the right of Figure \ref{visenco}. The encoding result is  $\hat{V}=\left[\hat{v}_{1} ; \hat{v}_{2} ; \hat{v}_{3}; \ldots \hat{v}_m\right]$, where $\hat{V} \in \mathbb{R}^{m \times d}$:
\begin{eqnarray}
\hat{v}_i = \frac{m}{n}\sum_{j\in Group_i} M(v_j), \text{  }i=1 \text{ to } m.
\end{eqnarray}

This encoded visibility is then used as the input of the cross-attention module in the U-Net.

\subsubsection{Cross-Attention Fusion} 
Next, we employ a cross-attention mechanism \cite{vaswani2017attention, rombach2022high} to fuse the encoded visibility information into the U-Net. Cross-attention can effectively learn relationships between different data domains. Specifically, we use the concatenation of both image features and visibility features to form $\textit{Key}$ and $\textit{Value}$:
\begin{eqnarray}
{CrossAttention}  = \operatorname{softmax}\left(\frac{\textit{Query} \cdot \textit{Key} ^{T}}{\sqrt{d}}\right) \cdot \textit{Value},
\end{eqnarray}
where
\begin{eqnarray}
\textit{Query}  =  W_{\textit{Query}} \cdot \varphi\left(\mathbf{x}_t, t, \mathbf{x^d}\right), \\
\textit{Key}  =  W_{\textit{Key}} \cdot \operatorname{Concat}(\varphi\left(\mathbf{x}_t, t, \mathbf{x^d}\right), \text{E}(V)), \\
\textit{Value}  =  W_{\textit{Value}} \cdot \operatorname{Concat}(\varphi\left(\mathbf{x}_t, t, \mathbf{x^d}\right), \text{E}(V)).
\end{eqnarray}
Here, $\varphi\left(\mathbf{x}_t, t, \mathbf{x^d}\right)$ denotes the intermediate latent representation of the U-Net that implements the predictor $\epsilon_\theta$. $W_{\textit{query}}, W_{\textit{Key}}, W_{\textit{Value}}$ are learnable weights. $\text{E}(V)$ refers to the encoded visibility tokens.

The aim of this cross-attention mechanism is to weigh the contributions of features from both the dirty image and the sparse visibility, allowing the U-Net to selectively attend to the most relevant information when generating the reconstructed image. 

\hspace*{\fill}

In summary,  we design the $\boldsymbol{\epsilon}_{\theta}\left(\mathbf{x}_{t}, t, \mathbf{x^d}, V\right)$ as follows: 
$\boldsymbol{\epsilon}_{\theta}\left(\mathbf{x}_{t}, t, \mathbf{x^d}, V\right) = f_{out}(\text{Cross-Attn}(f_{in}(g(\mathbf{x}_{t}, \mathbf{x^d}), t),\text{E}(V)), t)$

where $f_{in}$ refers to the input blocks , and $f_{out}$ the output blocks after the cross-attention module in the U-Net structure, as illustrated in Figure \ref{unet}. $\text{E}(V)$ refers to the encoded visibility tokens, and $g(\cdot, \cdot)$ the concatenation of images by channel.

\section{Experiments}
\label{Experiments and Results}
In this section, we perform a comprehensive evaluation on our method in comparison with several classic and recent state-of-the-art methods to demonstrate the overall improvement achieved by our approach. We also conduct ablation experiments to study the effects of individual techniques in our method. Additionally, we vary the number of sample noise used to obtain a result clean image, the number of diffusion steps to perform towards each noise image, and the total number of training images to examine our model performance.

\subsection{Experimental Setup} 
\textbf{Platform}. We conduct all experiments on a server with two AMD EPYC 7302 CPUs, 128GB main memory, and eight Nvidia RTX 3090 GPUs each with 24GB device memory. The server is equipped with an NVME 2TB SSD and four 1TB SATA hard disks. The operating system is Ubuntu 20.04. Our model is implemented in PyTorch 1.8.1 \cite{paszke2019pytorch}.

\textbf{Datasets}. We use the Galaxy10 DECals dataset \cite{Galaxy10} in our experiments, as in the most recent work on astronomical interferometric image reconstruction \cite{wu2022neural}. This dataset contains a total of 17,736 galaxy images, derived from the DESI Legacy Imaging Surveys \cite{dey2019overview}, which combine data from the Beijing-Arizona Sky Survey (BASS) \cite{zou2017project}, the DECam Legacy Survey (DECaLS) \cite{blum2016decam}, and the Mayall z-band Legacy Survey \cite{silva2016mayall}. In our experiments, all images are of a size 128 $\times$ 128 in pixels. With these images as the ground truth, we then use the eht-imaging toolkit \cite{chael2019ehtim, chael2018interferometric} to generate visibility data in the form of $\left\{u_{s}, v_{s}, {V}\left(u_{s}, v_{s}\right)\right\}$, with the observation parameters set to match an 8-telescope Event Horizon Telescope (EHT) array \cite{wu2022neural}.  The number of sampled visibility points on each image is 1660. Also following Wu et al. \cite{wu2022neural}, we use the discrete Fourier transform (DFT) method to generate dirty images from the visibility data. Finally, we randomly split the 17,736 images into a training set of 17,224 images and a testing set of 512 images following Wu et al. \cite{wu2022neural}.

\textbf{Implementation Details}.
Our experimental model architectures are based on U-Net \cite{ronneberger2015u} and augmented with timestep embedding and channel-wise self-attention block following DDPM \cite{ho2020denoising} and Guided-Diffusion \cite{dhariwal2021diffusion}. In visibility encoding, the number of elements in a positional embedding vector is set to 22 and the number of points in each point group is 20. In addition, we add spatial-wise attention layers and cross-attention layers in U-Net. Finally, the training batch size in VIC-DDPM is set to 32 images, and the total number of training steps is 25k. 

\textbf{Methods under Comparison}.
We compare VIC-DDPM with three other methods for radio interferometric image reconstruction, including the classic method CLEAN \cite{hogbom1974aperture} and recent deep learning-based approaches radionets \cite{schmidt2022deep} and Transformer-Conditioned Neural Fields \cite{wu2022neural}. We use the original code of each of these methods and follow the parameter setting in the original code for the best performance. In addition, we test our U-Net in both visibility domain only, U-Net(Vis), and image domain only, U-Net(Img), as supplementary baselines. 

\textbf{Evaluation Metrics}. 
We use two common metrics to evaluate image quality: Peak Signal-to-Noise Ratio (PSNR) and Structural Similarity Index Measure (SSIM). They quantify the overall image quality of reconstructed images and the perceptual similarity to the ground truth images, respectively. We compute these two metrics using the scikit-image package \cite{singh2019basics}, which follow the formulas presented by Hore et al. \cite{hore2010image}.

\subsection{Overall Comparison}

\begin{figure*}[t]
\centerline{\includegraphics[width=14cm]{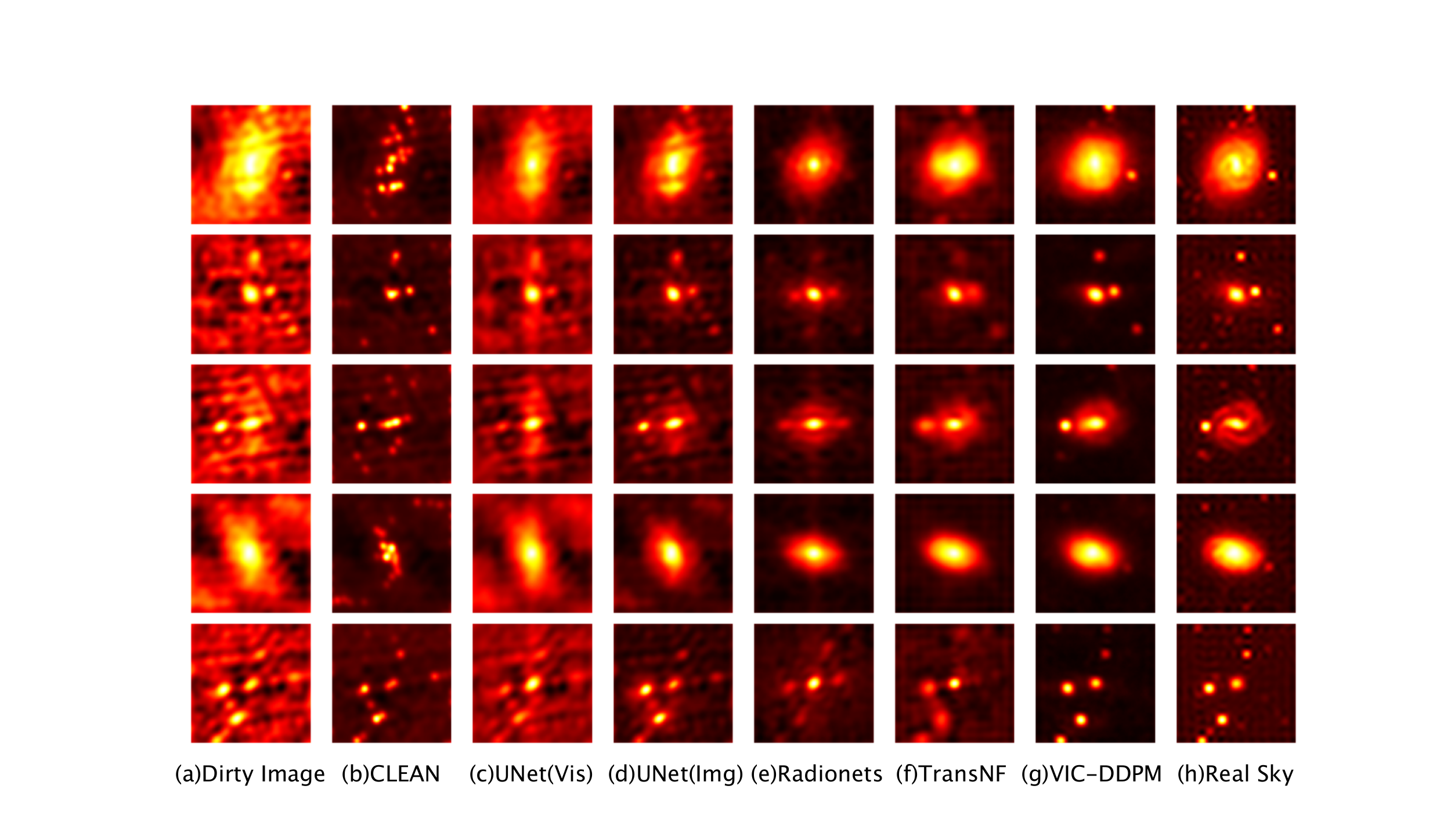}}
\caption{Visual Results of Overall Comparison.}
\label{Visual}
\end{figure*}

First, we present in Figure \ref{Visual} five representative reconstructed images on the Galaxy10(DECals) dataset from all methods under comparison as well as the dirty images and the ground truth images. 
Comparing the dirty images in Figure \ref{Visual} (a) and the ground truth in Figure \ref{Visual} (h), we see that dirty images contain many artifacts and that actual objects are often distorted. 

As shown in Figure \ref{Visual} (b), the CLEAN \cite{hogbom1974aperture} method reduces most artifacts, but misses the original structures of the sources. Due to the assumptions in the CLEAN \cite{hogbom1974aperture} algorithm, the emission distribution cannot be smoothly reconstructed. In comparison, the U-Net approach in the visibility domain (Figure \ref{Visual} (c)) leaves a lot of artifacts. The spatial domain U-Net (Figure \ref{Visual} (d)) achieves relatively better reconstruction, but does not fix the distortion and overlooks some faint sources or generates non-existent faint sources. Radionets \cite{schmidt2022deep} (Figure \ref{Visual} (e)) performs well in denoising and recovering from distortion; however, it has problems distinguishing separate sources that are close to one another. The Transformer-Conditioned Neural Fields method (denoted TransNF) \cite{wu2022neural} (Figure \ref{Visual} (f)) faithfully recovers the main objects in the observation area except when the edges of the objects are blurry or when the surrounding objects are small and dim.

The results of VIC-DDPM (Figure \ref{Visual} (g)) show that our method effectively removes artifacts and reconstructs details of the true objects, including relatively small or faint sources and detailed structures. These results demonstrate the effectiveness of our DDPM conditioned with both dirty images and visibility data.  

Next, we compute the PSNR and SSIM scores, with mean and standard deviation, for all 512 reconstructed images produced by our method and the methods under comparison. The results are shown in Table \ref{table_1}. The greater the PSNR and SSIM scores, the better.

The experimental results demonstrate that our method consistently achieves better performance in both PSNR and SSIM, indicating the effectiveness of our approach in providing accurate reconstructions.

\begin{table}
  \centering
  \caption{Performance of Different Methods.}
    \begin{tabular}{c | c c | c c}
    \toprule
    \multirow{2}{*}{Models}  & \multicolumn{2}{c|}{Domain} & \multicolumn{2}{c}{Metrics} \\
    \cmidrule{2-3} \cmidrule{4-5}
          & Img  & Vis  & PSNR$\uparrow$  & SSIM$\uparrow$ \\
    \midrule
    Dirty Image & &  & 10.406 $\pm$ 1.101 & 0.6321 $\pm$ 0.0590  \\
    \midrule
    CLEAN  & \checkmark &  & 17.366 $\pm$ 2.406 & 0.7130 $\pm$ 0.0405 \\
    \midrule
    U-Net (Img)  & \checkmark &   & 21.414 $\pm$ 1.654 & 0.8028 $\pm$ 0.0194 \\
    \midrule
    U-Net (Vis)  &  & \checkmark  & 13.015 $\pm$ 1.103 &  0.7079 $\pm$ 0.0230 \\
    \midrule
    Radionets  &  & \checkmark  &  21.433 $\pm$ 2.105 & 0.7940 $\pm$ 0.0237 \\
    \midrule
    Transformer\_NF & &  \checkmark & 20.707 $\pm$ 2.970 & 0.8084  $\pm$ 0.0404 \\
    \midrule
    \textbf{VIC-DDPM} &\checkmark & \checkmark & \textbf{22.615 $\pm$ 1.531} & \textbf{0.8242 $\pm$ 0.0224} \\
    \bottomrule
    \end{tabular}
  \label{table_1}
\end{table}

\subsection{Ablation Experiments}

In the ablation experiments, we test the following variants of our method by selectively removing one or more components to investigate the individual contributions of the components to the performance: (1) without any conditioning; (2) without the visibility data conditioning; and (3) without the dirty image conditioning.
We train these variants by add no conditioning, replacing the visibility data with zeros, and by replacing the dirty image with zeros, respectively. We compare these three variants with the original dirty images and the reconstructed images from the full VIC-DDPM.

\begin{figure}
\centerline{\includegraphics[width=10cm]{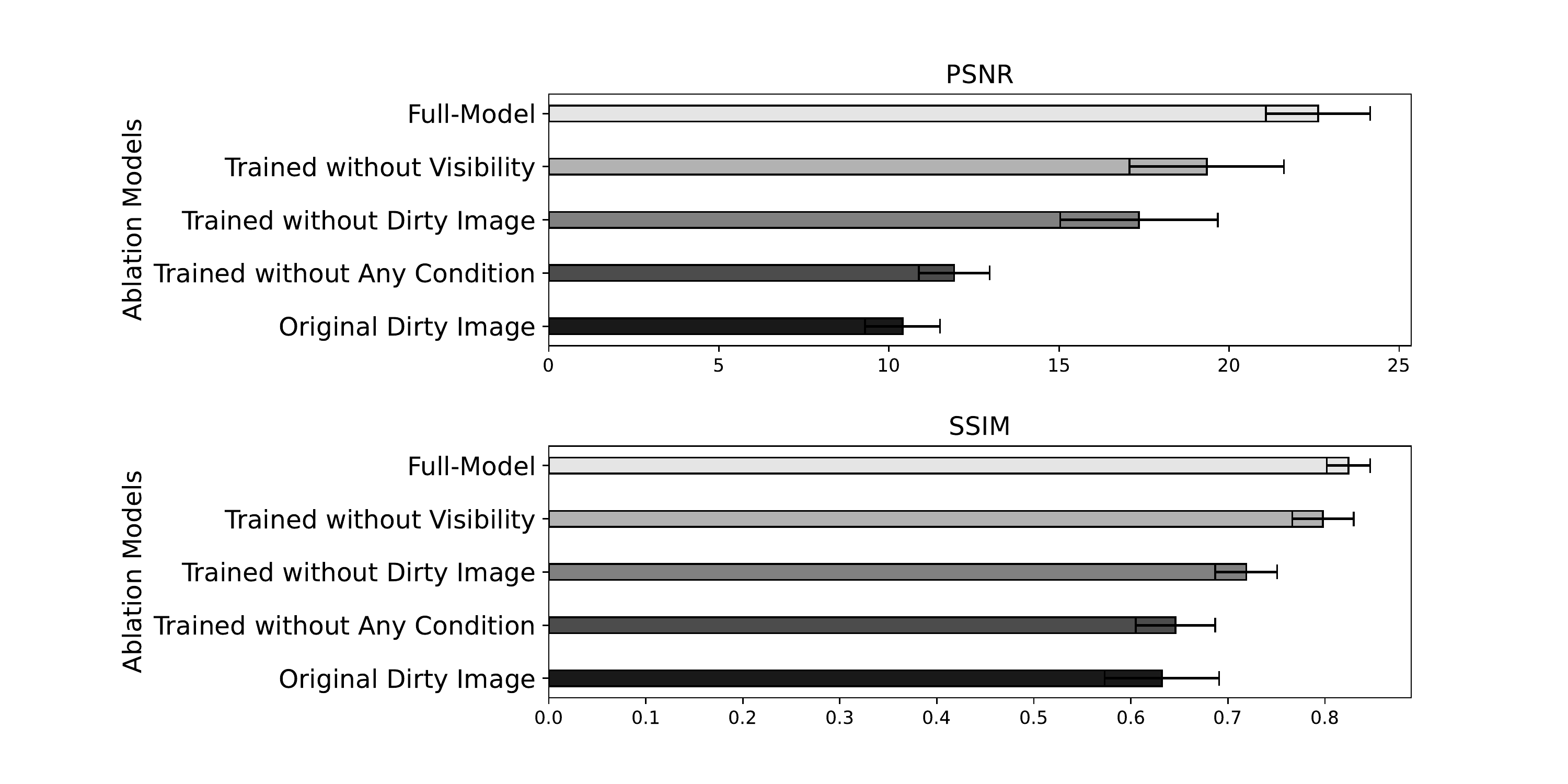}}
\caption{Ablation Results.}
\label{ablation}
\end{figure}

\begin{figure}
\centerline{\includegraphics[width=9cm]{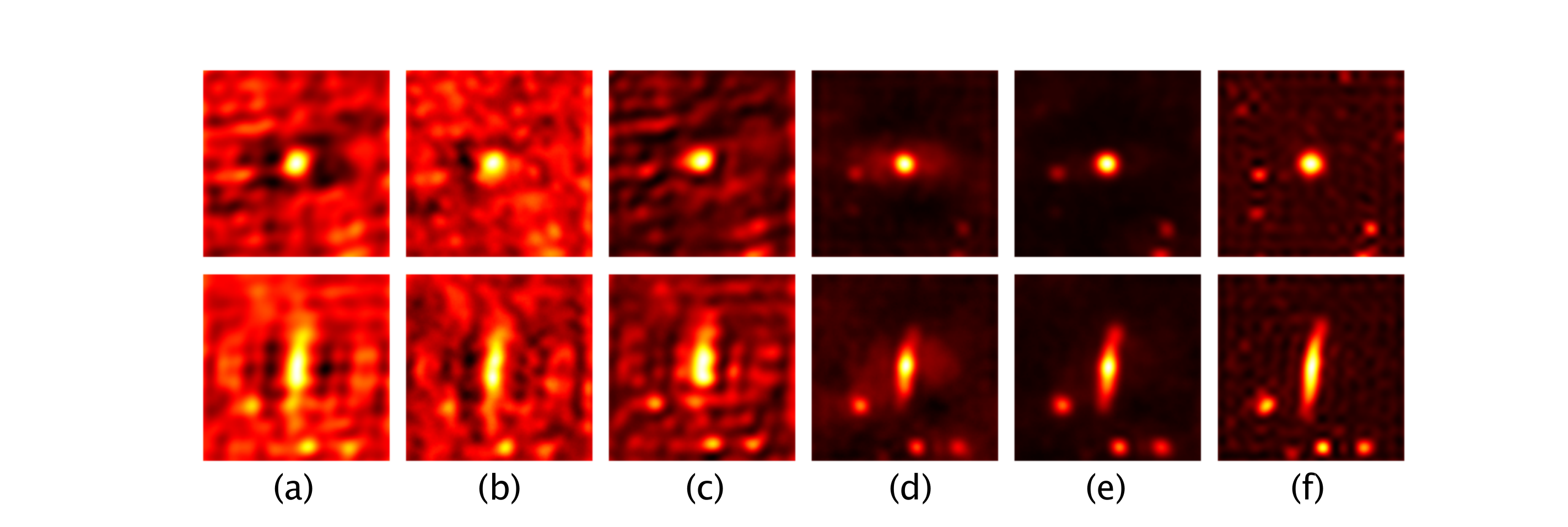}}
\caption{Visual Examples of Ablation Results. Columns from let to right are (a) original dirty image, (b) unconditional DDPM, (c) our model trained without dirty image conditioning, (d) our model trained without visibility conditioning, (e) full model of VIC-DDPM, and (f) real sky, respectively. }
\label{ablation_visual}
\end{figure}

We evaluate the performance of each ablation model using the same dataset and evaluation metrics (PSNR and SSIM) as in the previous experiment. The results in Figure \ref{ablation} show that our full VIC-DDPM model consistently achieves the best performance, demonstrating the effectiveness of combining dirty image conditioning and visibility conditioning. The dirty image conditioning seems more effective than the visibility data conditioning, suggesting that spatial information plays a more important role than signal-to-noise separation in DDPM. Some visual examples of ablation results are shown in Figure \ref{ablation_visual}. These results confirm that the full model (column e) is best, and that dirty image conditioning (column d) is more effective than visibility conditioning (column c). Without any conditioning, column (b) is the least effective.

\subsection{Effect of Parameter Setting}

\begin{figure}[t]
    \centering
    \subfigure[]
    {\includegraphics[height=3.8cm, width=4.2cm]{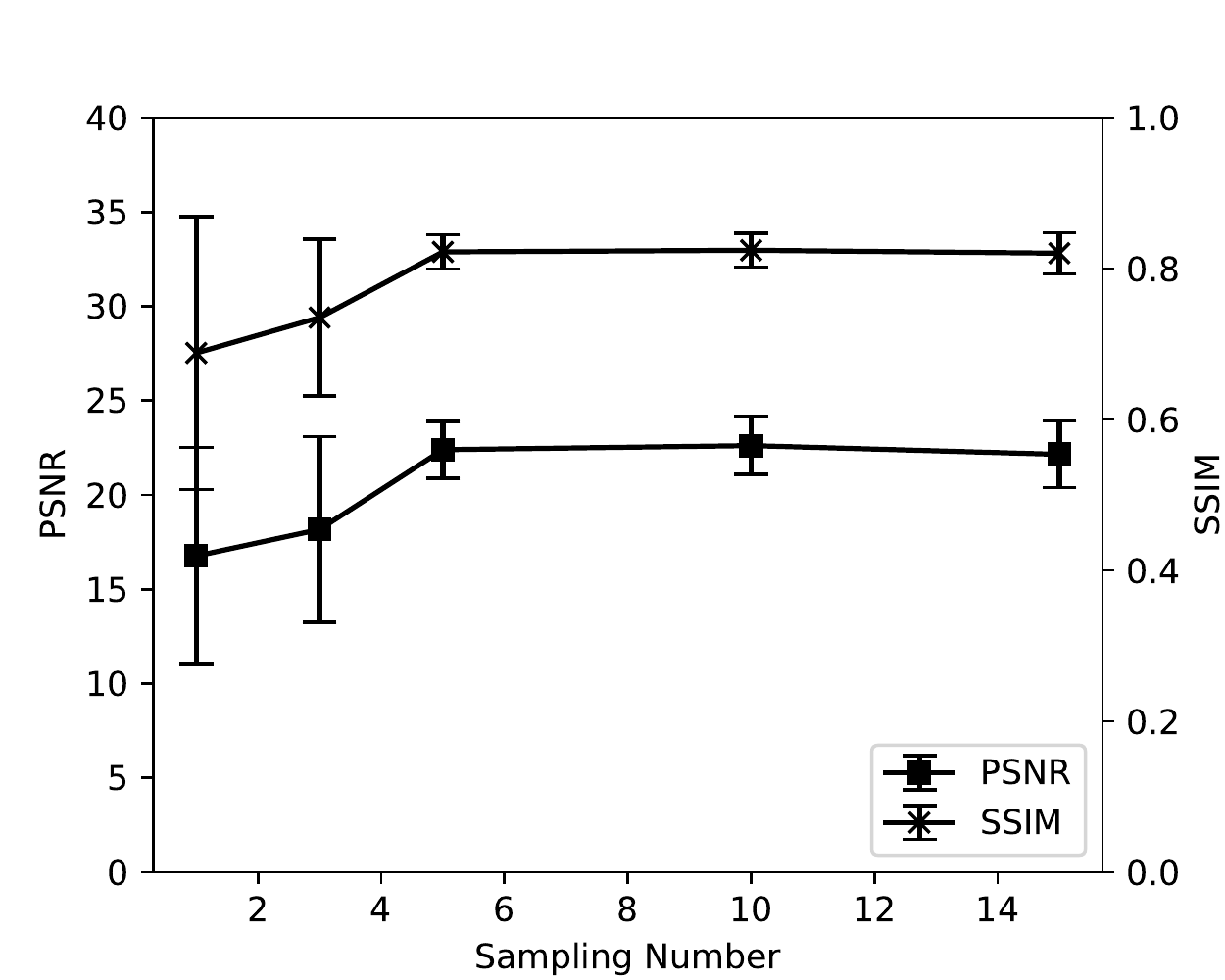}}
    \subfigure[]
    {\includegraphics[height=3.8cm, width=4.2cm]{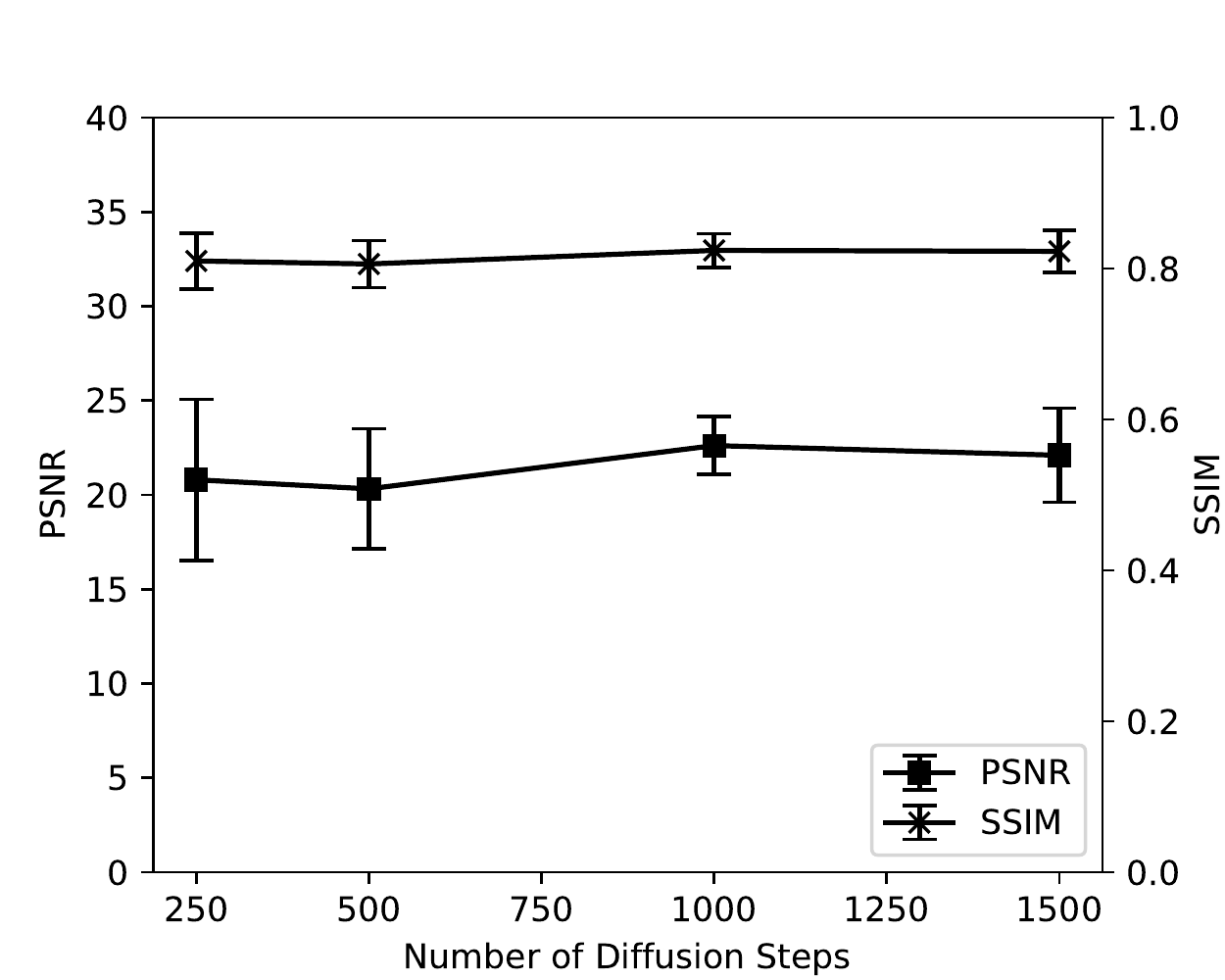}}
    \caption{Effect of Parameters: (a) number of random noise images to test for each dirty image; (b) number of diffusion steps. }
    \label{tradeoff}
\end{figure}

We also conduct experiments to investigate the effect of two important parameters in our model --  number of random noise sample images to test for each dirty image and number of diffusion steps. We report the performance on PSNR and SSIM in Figure \ref{tradeoff}. Figure \ref{tradeoff}(a) shows that as the number of random noise sample images used increases for each test dirty image, the mean quality of the result images also increases; however, when the number of sample images further increases, the mean quality remains stable or slightly decreases. In comparison, Figure \ref{tradeoff}(b) shows that the result quality is almost constant regardless of number of diffusion steps. These results indicate that in our experimental setting, selecting 5 samples for each test dirty image and running each test for 1000 steps would be a good choice for best result quality.

\subsection{Effect of Training Data Size}
\begin{table}
  \centering
  \caption{Performance (PSNR/SSIM) with training dataset size varied}
    \begin{tabular}{c|c|c|c}
    \toprule
    \multirow{2}{*}{Data} & \multicolumn{3}{c}{Methods} \\
    \cmidrule{2-4}
          & RadioNet & Transformer\_NF & VIC-DDPM \\
    \midrule
    \multirow{2}{*}{0.5k}  & 19.401 $\pm$ 1.639 & 18.235 $\pm$ 2.802 & 17.149 $\pm$ 6.856 \\
         & 0.7586 $\pm$ 0.0210 & 0.7610 $\pm$ 0.0634 & 0.7417 $\pm$ 0.1220 \\
    \midrule
    \multirow{2}{*}{1k} & 20.267 $\pm$ 1.822 & 18.935 $\pm$ 2.900 & 22.420 $\pm$ 3.765 \\
         & 0.770 $\pm$ 0.0222 & 0.7781 $\pm$ 0.0675 & 0.8236 $\pm$ 0.0354 \\
    \midrule
    \multirow{2}{*}{2k} & 20.610 $\pm$ 1.926 & 18.565 $\pm$ 2.360 & 22.170 $\pm$ 1.986 \\
         & 0.7765 $\pm$ 0.0241 & 0.7735 $\pm$ 0.0578 & 0.8137 $\pm$ 0.0211 \\
    \midrule
    \multirow{2}{*}{4k} & 20.657 $\pm$ 1.996 & 20.069 $\pm$ 2.611 & 22.403 $\pm$ 2.172 \\
         & 0.7787  $\pm$ 0.0269 & 0.8028 $\pm$ 0.0518 & 0.8210 $\pm$ 0.0328 \\
    \midrule
    \multirow{2}{*}{17k} & 21.433 $\pm$ 2.105 & 20.707 $\pm$ 2.970 & 22.615 $\pm$  1.531 \\
         & 0.7940 $\pm$ 0.023 & 0.8084 $\pm$ 0.040 & 0.8242 $\pm$ 0.022 \\
    \bottomrule
    \end{tabular}
  \label{dataeff_table}
\end{table}

Model performance may vary with the training dataset size. We therefore use different sizes of training datasets to evaluate our model performance in PSNR/SSIM in comparison with two latest deep-learning methods -- Radionets \cite{schmidt2022deep} and Transformer-Conditioned Neural Fields (denoted Transformer\_NF) \cite{wu2022neural}. To create training sets of different sizes, we randomly choose 500, 1000, 2000, and 4000 of the 17k training images in the Galaxy10 DECals dataset \cite{Galaxy10}. As shown in Table \ref{dataeff_table}, when the dataset size increases from 0.5k to 17k, the PSNR/SSIM performance increases by 10.5\%/4.7\%, 13.6\%/6.2\% and 31.9\%/11.1\% on three models, respectively. Compared with the other two methods, the performance of VIC-DDPM significantly improves when the dataset size increases from 500 to 1000. With a training set of 1000 images, VIC-DDPM performance is already considerably higher than the other two methods,  and further increase of training dataset size shows little effect. 
In contrast, both radionets and Transformer\_NF continue to improve as the training dataset size increases, but the performance improvement is not as significant as VIC-DDPM. As a result, VIC-DDPM outperforms the other two on data sizes from 1K to 17K. This result suggests that VIC-DDPM requires a much smaller training set than the other two to obtain high-quality reconstruction. We plan to find larger training datasets to give more room to the other two methods for performance improvement.

\begin{figure}[t]
    \centering
    \subfigure[]
    {\includegraphics[width=4.25cm]{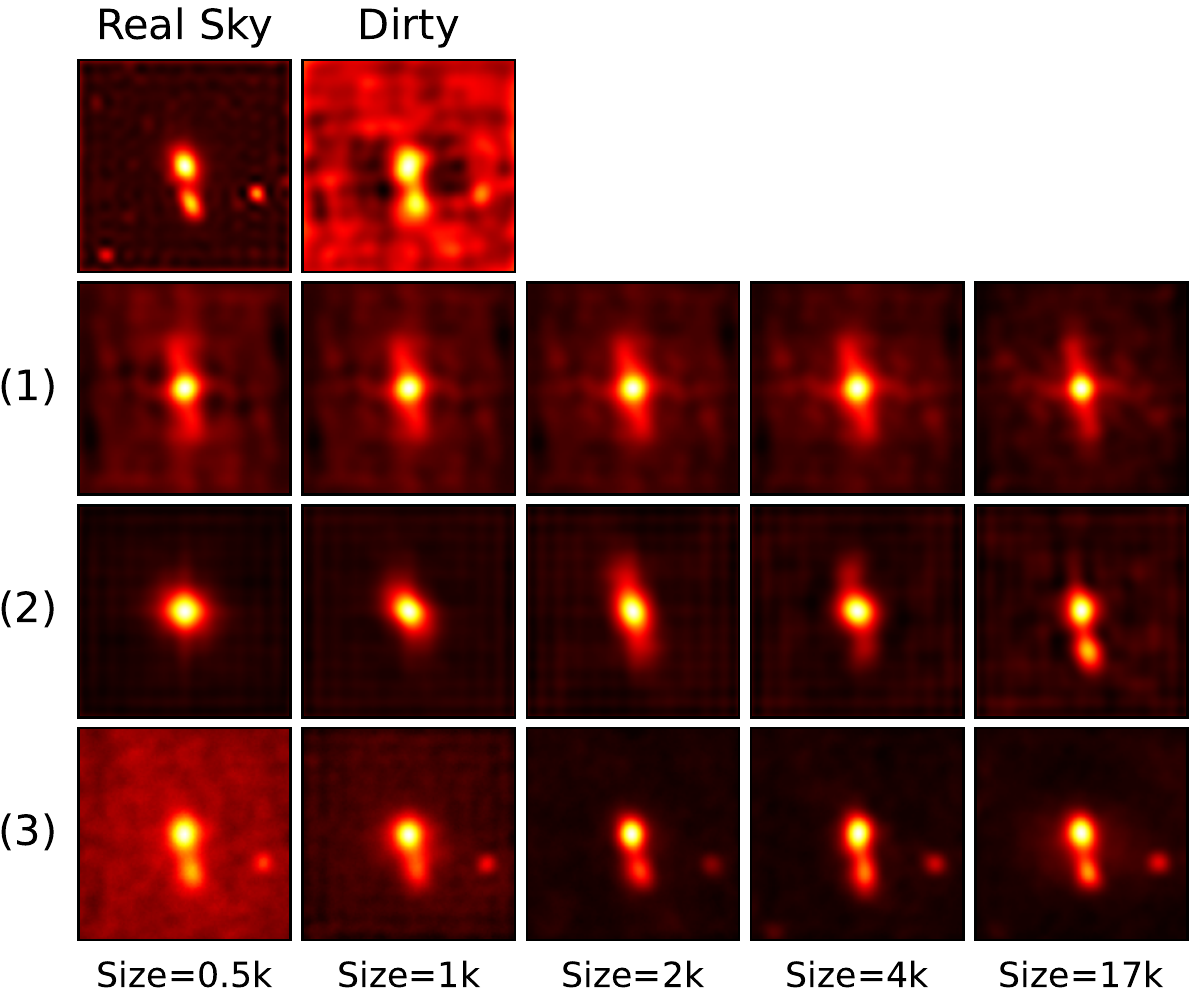}}
    \subfigure[]
    {\includegraphics[width=4.25cm]{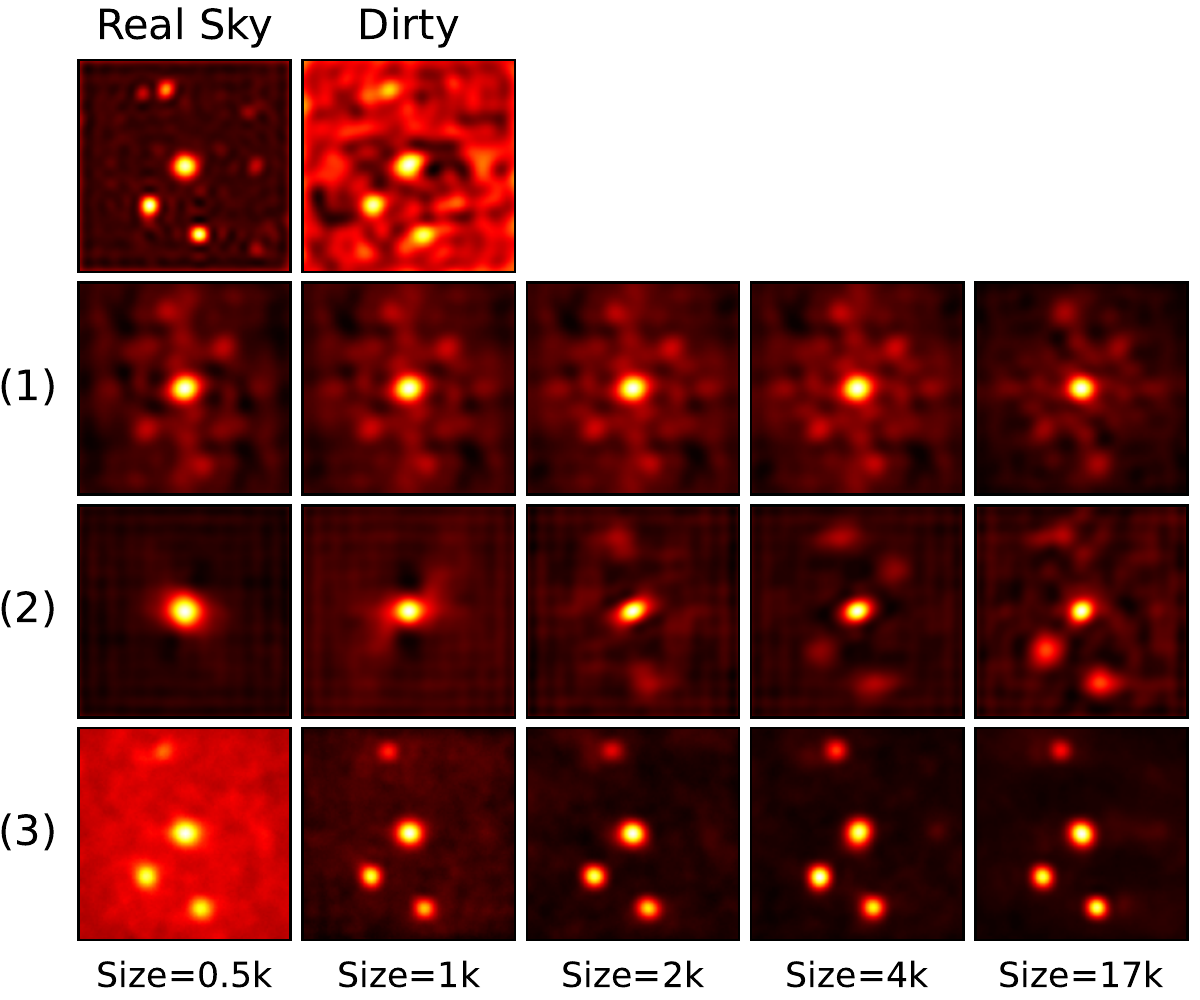}}
    \\
    \subfigure[]
    {\includegraphics[width=4.25cm]{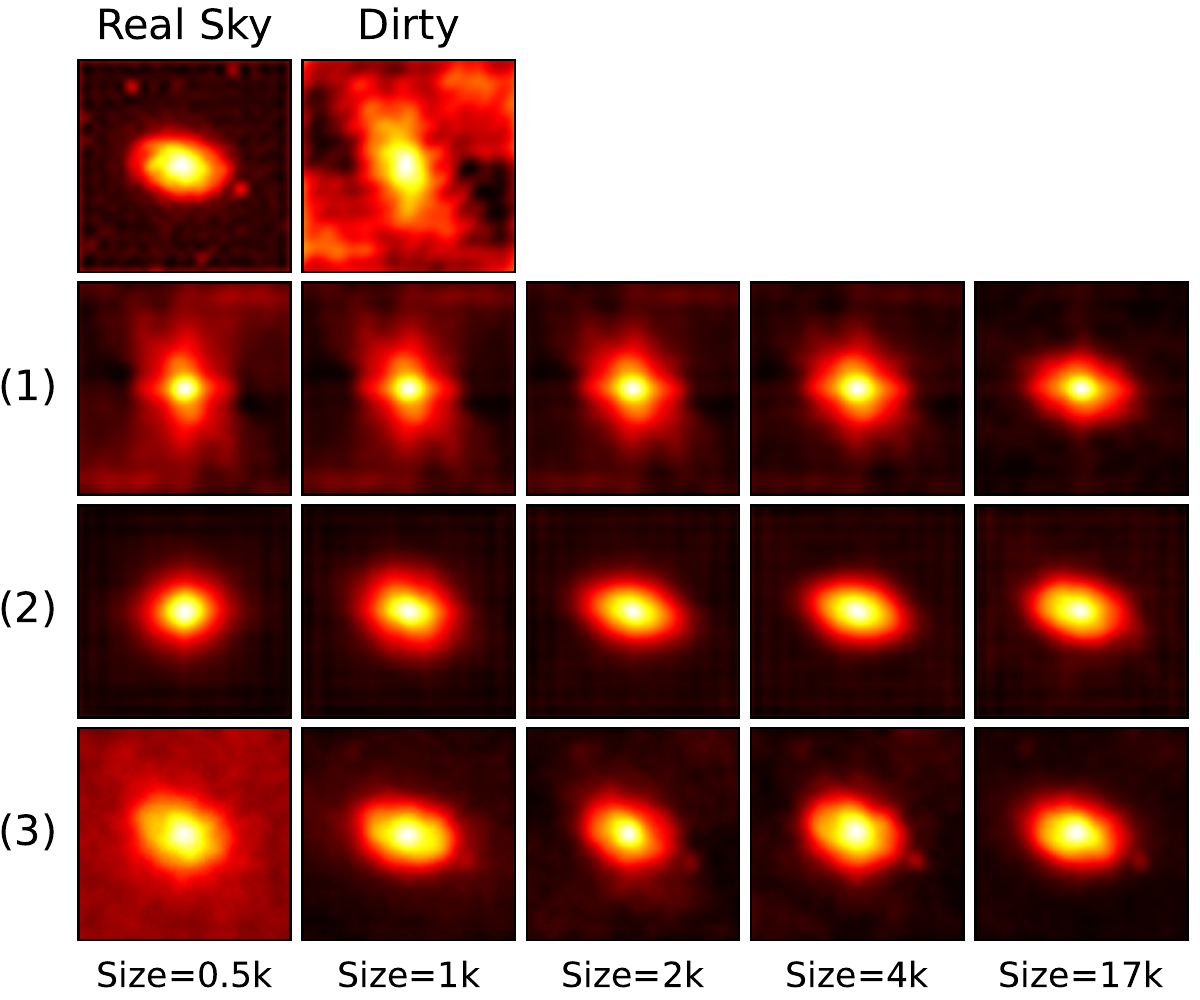}}
    \subfigure[]
    {\includegraphics[width=4.25cm]{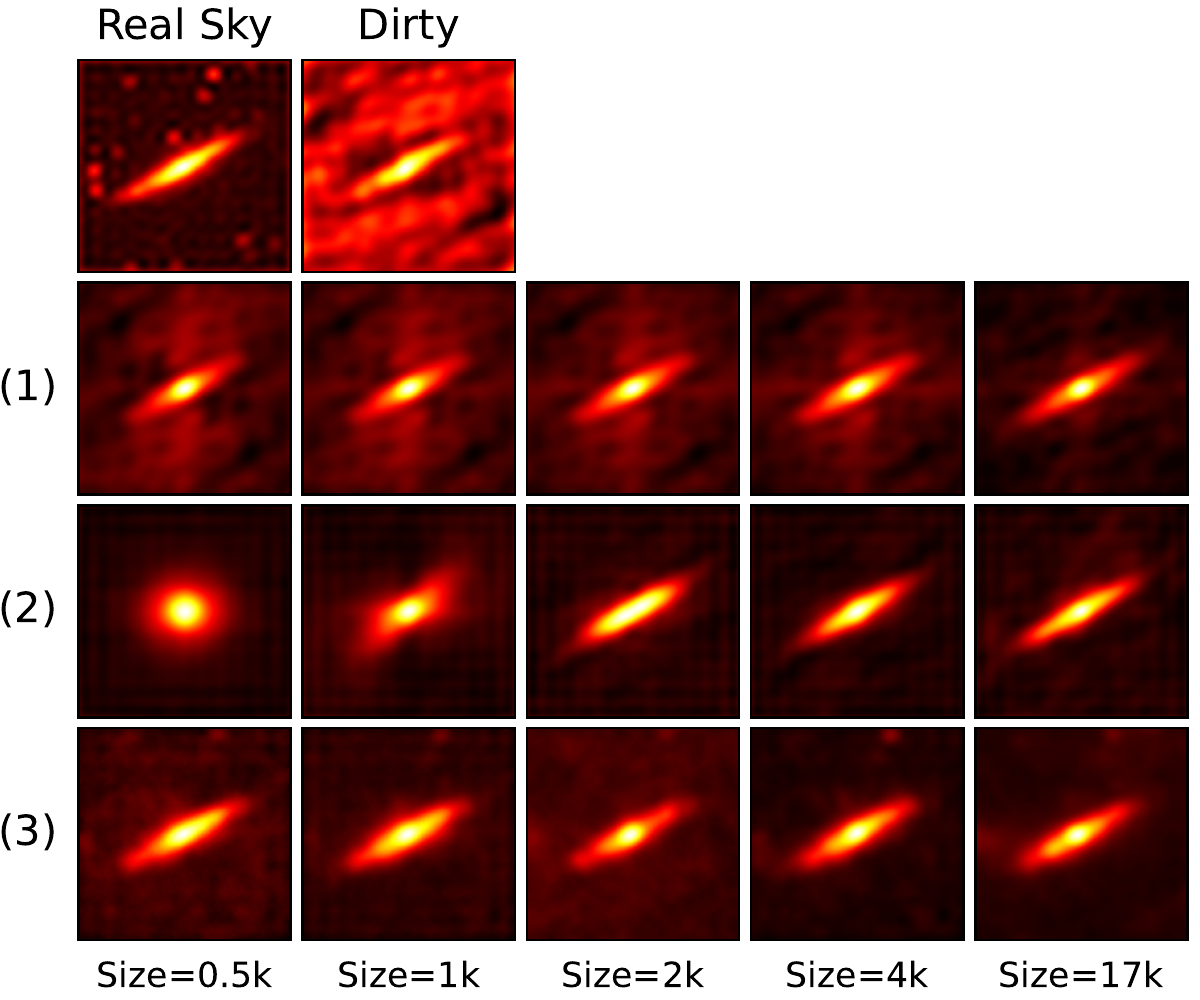}}
    \caption{Visual results on four example images (a)-(d) from three models including (1) radionets, (2) Transformer-Conditioned Neural Fields, and (3) VIC-DDPM trained with datasets of different sizes.}
    \label{dataeff_egs}
\end{figure}

To further investigate the performance differences, we examine the visual effects of reconstructed images and provide a few examples with different training dataset sizes in Figure \ref{dataeff_egs}. Subfigures on each row are from a single method, and those on each column are on the same training dataset size. We find that, even though our model is slightly lower than the other two methods on the numeric performance metrics when the training set size is 500 in Table \ref{dataeff_table}, our method's reconstruction of the structures, in the bottom left of each example (a) through (d) is noticeably superior to the other two methods when visually inspected by the human eye, as shown in Figure \ref{dataeff_egs}. These visual results suggest that the lower metric values of our method at 500 training images are possibly because the background noise has not been effectively reduced by then.

 As observed in Figure \ref{dataeff_egs} (a) , (c) and (d), VIC-DDPM is capable of reconstructing the basic contours of astronomical sources after learning only 500 samples. In contrast, other methods require learning from 1000 to 4000 samples to reconstruct the basic contours. On reconstruction of images containing multiple astronomical sources, as shown in Figure \ref{dataeff_egs} (a) and (b), our method can discern the positions and structures of multiple astronomical sources after learning from 500 training samples, while other methods cannot.

\section{Conclusion and Future Work}
In conclusion, we have presented VIC-DDPM, a Visibility and Image Conditioned Denoising Diffusion Probabilistic Model, for radio interferometric image reconstruction. Inheriting the capability of fine detail generation from DDPM as well as incorporating the complementary strengths of data in spectral and spatial domains, VIC-DDPM improves the accuracy of reconstructed images with effective removal of artifacts, and recovery of structural details and dim sources. Experimental results demonstrate the superiority of VIC-DDPM over both traditional and deep-learning based approaches on the Galaxy10 DECals dataset. 

In the future, we will apply our methods on both real and synthetic datasets based on different kinds of radio telescope arrays and generalize our methods to images in other domains.

\ack This work was partially supported by Grant 2020SKA0110300 from the National SKA Program of China.

\bibliography{ecai}

\end{document}